\documentclass[conference]{IEEEtran}
\IEEEoverridecommandlockouts
\usepackage{cite}
\usepackage{amsmath,amssymb,amsfonts}

\usepackage{algorithmic}
\usepackage{graphicx}
\usepackage{textcomp}
\usepackage{xcolor}
\usepackage{balance}

\usepackage{breqn}
\usepackage[font=footnotesize,skip=2pt]{caption}
\ifCLASSOPTIONcompsoc
    \usepackage[caption=false, font=normalsize, labelfont=sf, textfont=sf]{subfig}
\else
\usepackage[caption=false, font=footnotesize]{subfig}
\fi

\newenvironment{nohyphens}{%
  \par
  \hyphenpenalty=100000
  \exhyphenpenalty=100000
  \sloppy % Makes TeX obey margins by stretching inter word spaces
}{\par}

\usepackage[linesnumbered,ruled,vlined]{algorithm2e}

\SetCommentSty{mycommfont}
\SetKwInput{KwInput}{Input}                % Set the Input
\SetKwInput{KwOutput}{Output}

\def\BibTeX{{\rm B\kern-.05em{\sc i\kern-.025em b}\kern-.08em
    T\kern-.1667em\lower.7ex\hbox{E}\kern-.125emX}}
\begin{document}

\title{FastCaps: A Design Methodology for Accelerating Capsule Network on Field Programmable Gate Arrays\vspace{-10pt}}

\author{\IEEEauthorblockN{ Abdul Rahoof}
\IEEEauthorblockA{ 
\textit{Indian Institute of Technology Palakkad}\\
India \\
112004001@smail.iitpkd.ac.in}
\and
\IEEEauthorblockN{ Vivek Chaturvedi}
\IEEEauthorblockA{ 
\textit{Indian Institute of Technology Palakkad}\\
India \\
vivek@iitpkd.ac.in}
\and
\IEEEauthorblockN{ Muhammad Shafique}
\IEEEauthorblockA{
\textit{New York University}\\
Abu Dhabi, UAE \\
muhammad.shafique@nyu.edu}

}

\maketitle

\begin{abstract}
Capsule Network (CapsNet) has shown significant improvement in understanding the variation in images along with better generalization ability compared to traditional Convolutional Neural Network (CNN). CapsNet preserves spatial relationship among extracted features and apply dynamic routing to efficiently learn the internal connections between capsules. However, due to the capsule structure and the complexity of the routing mechanism, it is non-trivial to accelerate CapsNet performance in its original form on Field Programmable Gate Array (FPGA). Most of the existing works on CapsNet have achieved limited acceleration as they implement only the dynamic routing algorithm on FPGA, while considering all the processing steps synergistically is important for real-world applications of Capsule Networks. Towards this, we propose a novel two-step approach that deploys a full-fledged CapsNet on FPGA. First, we prune the network using a novel Look-Ahead Kernel Pruning (LAKP) methodology that uses the sum of look-ahead scores of the model parameters. Next, we simplify the nonlinear operations, reorder loops, and parallelize operations of the routing algorithm to reduce CapsNet hardware complexity. To the best of our knowledge, this is the first work accelerating a full-fledged CapsNet on FPGA. Experimental results on the MNIST and F-MNIST datasets (typical in Capsule Network community) show that the proposed LAKP approach achieves an effective compression rate of 99.26$\%$ and 98.84$\%$, and achieves a throughput of 82 FPS and 48 FPS on Xilinx PYNQ-Z1 FPGA, respectively. Furthermore, reducing the hardware complexity of the routing algorithm increases the throughput to 1351 FPS and 934 FPS respectively. As corroborated by our results, this work enables highly performance-efficient deployment of CapsNets on low-cost FPGA that are popular in modern edge devices.
\end{abstract}


\section{Introduction}
Convolution Neural Networks (CNNs) have become increasingly popular and have demonstrated good accuracy in various computer vision applications such as image classification~\cite{b1}, object detection~\cite{b2} and image segmentation~\cite{b3}. CNNs use a pooling layer between convolution layers to reduce the dimensions of feature maps and hence reduce the number of parameters to learn. Pooling layer passes the most important features from a region of a feature map to the next layer, thereby improve the  translational invariance property. However, this reduces the ability to recognize pose and object deformation as the relative spatial information of the features is lost in the process. Sabour et al. have proposed a Capsule Network (CapsNet), in which neurons are grouped together to form a layer of capsules and have removed the pooling layer by devising a powerful routing algorithm~\cite{b4}. The routing algorithm maps the capsule from a lower layer to an appropriate parent capsule in the layer above. Hence, CapsNet increases the ability to preserve spatial characteristics and improves the learning ability of the model, consequently increases the accuracy of the model. 
\par However, CapsNet has a higher MAC$\footnote{Multiplication and Accumulation}$/ Memory ratio and is more compute-intensive compared to CNNs, such as ResNet-18~\cite{b30} and AlexNet~\cite{b6}. Consequently, high computational efforts are needed to dynamically route the capsules resulting in a larger number of parameters and complex compute functions. Many of the use cases of CapsNet such as healthcare~\cite{b8} and autonomous vehicles~\cite{b9} require these networks to run on edge devices that have limited compute resources and work under severe energy constraints. Since CapsNet is a large and highly complex network, accelerating its performance on edge devices can be tremendously challenging. 
\par
\begin{figure}[h]
    \centering
  
  \subfloat[\label{1a}]{%
        \includegraphics[scale=0.1]{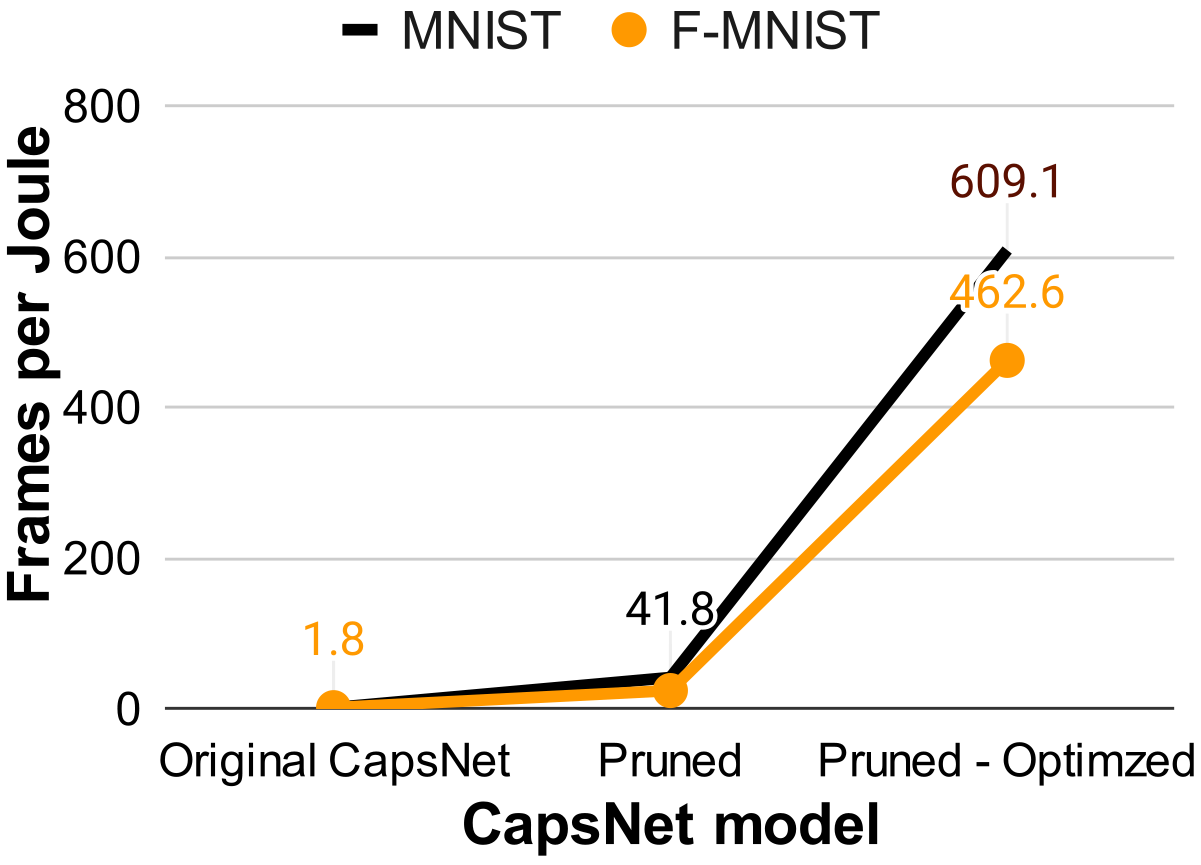}}
  \subfloat[\label{1b}]{%
        \includegraphics[scale=0.1]{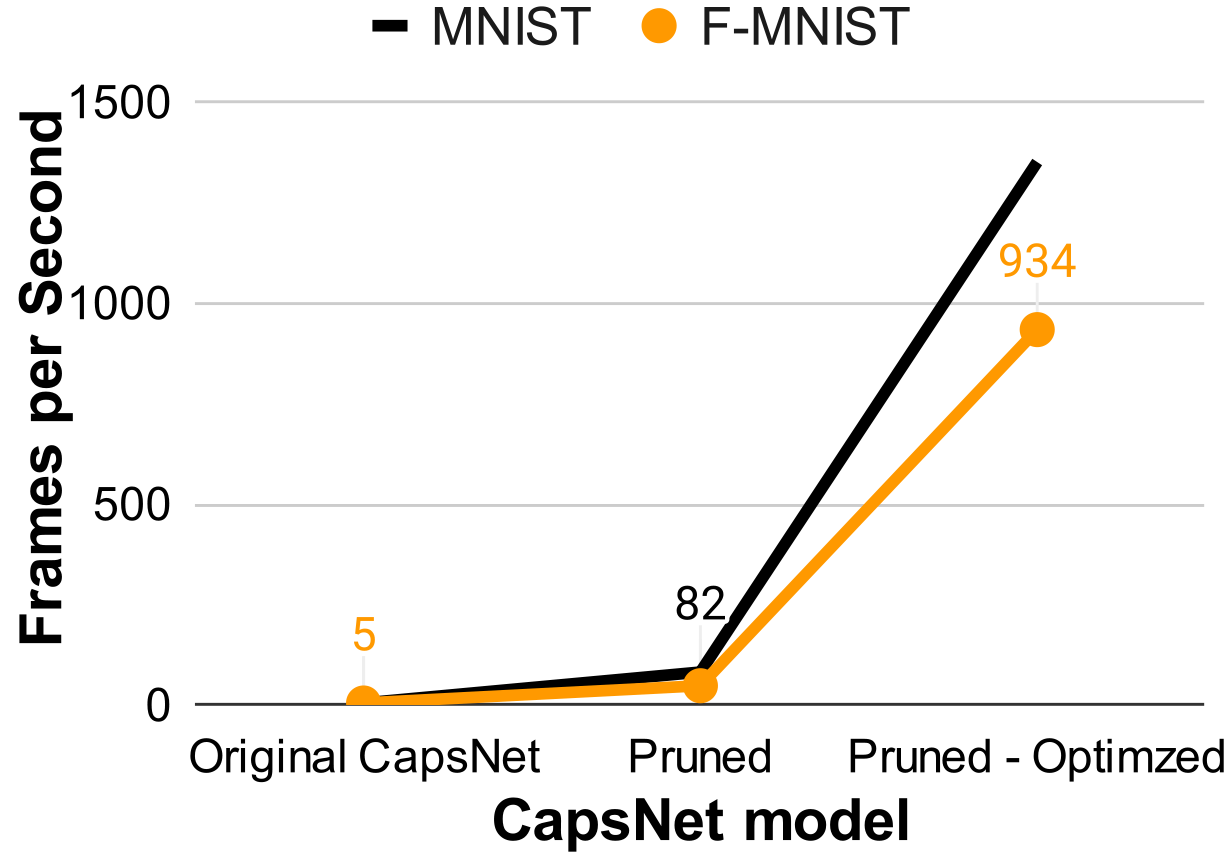}}
  %\vspace{-2mm}
  \caption{Energy utilization (a) and throughput (b) of the CapsNet model deployed on FPGA}
  \label{fig:my_label-10} 
\end{figure}

 Due to their energy efficiency, shorter time to market, and capability to be reprogrammed, FPGAs have become a popular acceleration platform integrated into modern edge devices. Unfortunately, most of the existing works on CapsNet are not designed to run on such devices~\cite{b7, b8, b9, b17, b18, b19, b32, b33}. Implementing CapsNet in its original form on an FPGA device is non-trivial due to the abundance of parameters, complex data flow of capsule processing and high MAC/Memory ratio. In general, to reduce the computation and memory of CNNs, various model compression techniques have been proposed, including quantization of the network parameters~\cite{b10}, low rank matrix approximations~\cite{b11}, circulant projection~\cite{b12} and pruning~\cite{b13,b14}. Particularly, pruning methods can provide higher compression rates and significantly reduce the redundancy in network parameters. Moreover, pruning can also improve computational latency and energy consumption when deployed on hardware devices~\cite{b14}. Fig.~\ref{fig:my_label-10} compares the throughput and energy consumption of the original, pruned, and pruned-optimized CapsNet model. The study indicates that pruning enhances performance, with the throughput rising from 5 FPS to 82 FPS and 48 FPS for MNIST and F-MNIST datasets, respectively. Additionally, energy efficiency improves from 1.8 FPJ to 41.8 FPJ and 24.5 FPJ respectively for the same datasets.   

\par In this paper, we propose a novel two-step approach for deploying the CapsNet model on FPGA. First, we perform pruning using our novel structured Look-Ahead Kernel Pruning (LAKP) technique, which prunes redundant kernels from a model based on the look-ahead score. The proposed LAKP is a simple pruning method and achieves better compression rates than the state-of-the-art magnitude-based kernel pruning~\cite{b14}. Note, the LAKP technique can be applied to compress any deep learning model that has convolution layers. In the case of CapsNet, pruning the kernel leads to a reduction in the number of capsules in the capsule layer, and a decrease in both the number of weight parameters and computation required for the routing algorithm. Experiments have shown that using LAKP with a pruning rate of $99.26\%$ resulted in a significant decrease in the number of weight parameters in the routing algorithm, up to 1280 times. Next, we reduce the hardware complexity of the dynamic routing algorithm by simplifying its non-linear operations, reordering the loops and parallelizing the operations.

\par
\begin{figure}[h]
\centering
\includegraphics[scale=0.14]{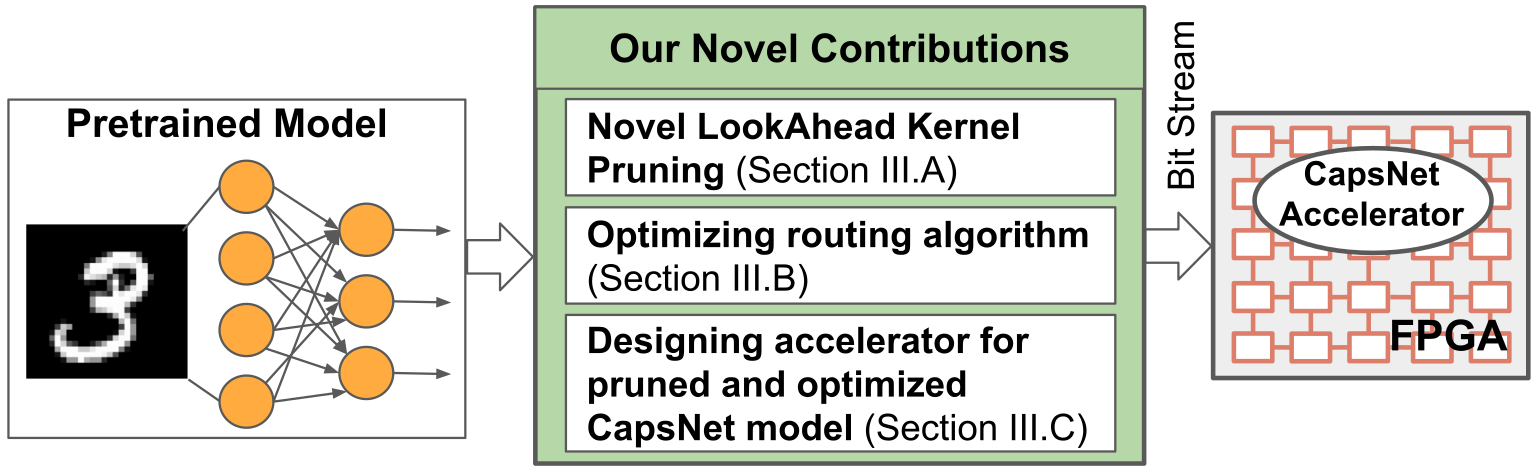}
%\vspace{-2mm}
\caption{Overview of our novel contributions}
\label{fig:my_label-0}
\end{figure}

 We used the Xilinx Vivado HLS platform for creating the overlay IP. To reduce the hardware complexity of the dynamic routing algorithm present in the CapsNet model, we expanded the non-linear functions such as exp() and div() using the Taylor series. Then, we re-ordered the loops and utilized the optimization directives of Vivado HLS to make the operations parallel. Experimental results on MNIST and F-MNIST datasets in Section~\ref{subsec-4b} demonstrate that the proposed CapsNet accelerator gained a significant throughput of 1351 FPS and 934 FPS respectively, which is 270x and 187x better than the original CapsNet implementation.
In Fig.~\ref{fig:my_label-0}, an overview of our novel approach to accelerate CapsNet on FPGA is demonstrated. 

\begin{nohyphens}
\par \textbf{In summary, our novel contributions are:}
\begin{itemize}
    \item We propose a structured kernel pruning method namely Look-Ahead Kernel Pruning (LAKP) which uses look-ahead scores of the parameters for pruning the kernels. Experimental results on CapsNet, ResNet-18 and VGG-19 models show that LAKP achieves consistently better compression rate compared to state-of-the-art magnitude-based kernel pruning (KP).
    \item We propose a novel strategy to simplify the dynamic routing algorithm by optimizing complex functions such as exp() and div() using the Taylor series. 
    \item We deployed both the original CapsNet and our proposed pruned and optimized CapsNet on Xilinx PYNQ-Z1 FPGA and achieved a considerable increase in performance, with 1351 FPS and 934 FPS respectively, providing 270x and 187x improvement over the original CapsNet implementation for MNIST and F-MNIST datasets respectively.
   
\end{itemize}
\end{nohyphens}

To the best of our knowledge we are the first to propose an FPGA-specific design optimization of CapsNet.

\section{Background and Related Work}
 The use of pooling layers in CNN breaks down the ability to count the spatial relationship of objects in an image that causes degradation of the performance when the training data is not sufficient. To address this challenge, Sabour et al. introduced a variant of CNN, called Capsule Network (CapsNet), which preserves the spatial relationship and improves the generalization ability of the model~\cite{b4}.     

\subsection{Capsule Network}
\par A capsule is a group of neurons organized in the form of a vector whose length represents the probability of occurrence of certain features and its individual dimensions represent various spatial characteristics such as width, thickness and rotation. The capsule structure and the powerful dynamic routing between the capsule layers replace the pooling layers used in CNNs. The dynamic routing algorithm maps the capsule from a lower layer to a particular capsule in a higher layer with the help of coupling coefficient. An important advantage of CapsNet is their ability to preserve spatial information of detected features when performing different types of recognition tasks. 
\par
\begin{figure}[h]
\centering
\includegraphics[scale=0.19]{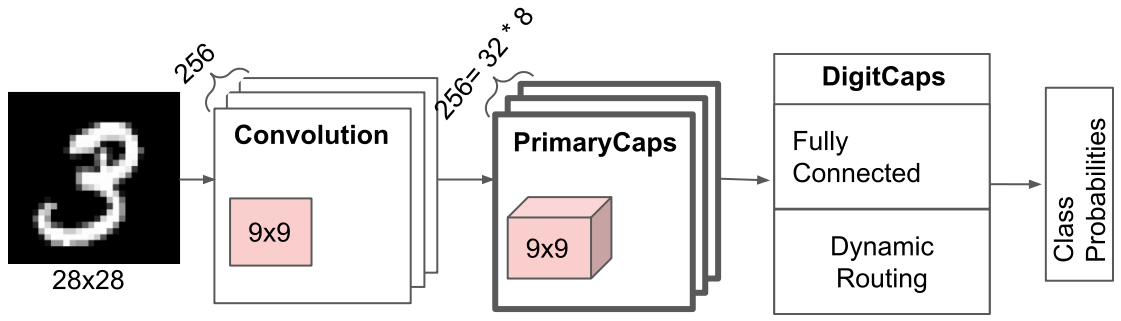}
%\vspace{-3mm}
\caption{CapsNet Architecture for inference \cite{b4}}
\label{fig:my_label-1}
\end{figure}

\begin{figure}
    \centering
    \includegraphics[scale=0.17]{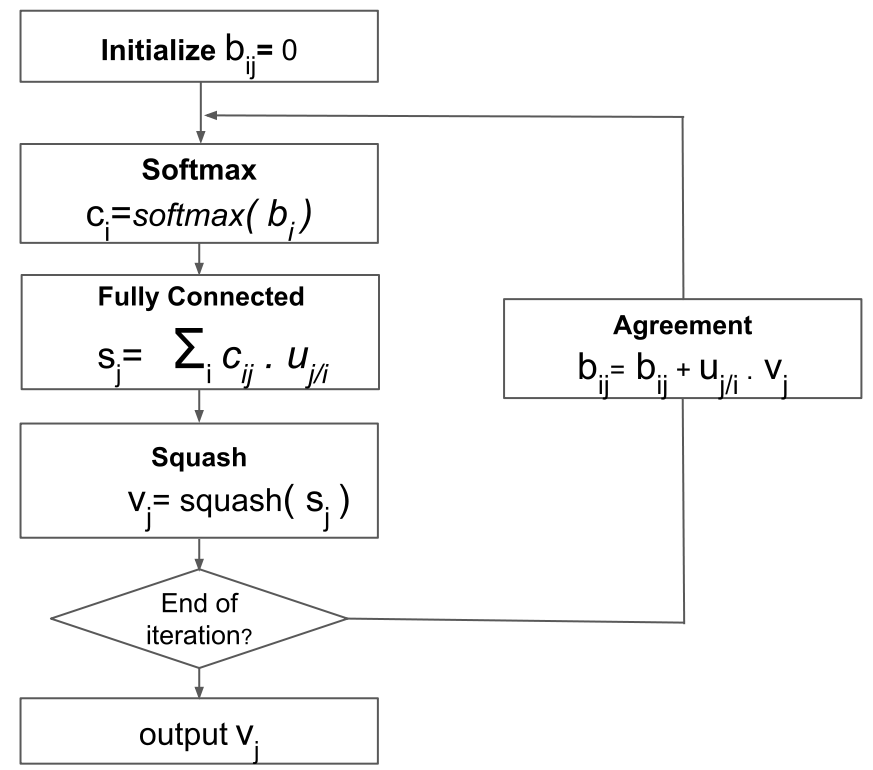}
    %\vspace{-3mm}
    \caption{Dynamic Routing Algorithm\cite{b4}}
    \label{fig:my_label-2}
\end{figure}

 The architecture of the CapsNet proposed in~\cite{b4} is shown in Fig.~\ref{fig:my_label-1}. It consists of the following three layers: Convolution (9x9 convolution/256 output channels), PrimaryCaps (9x9 convolution/256 output channels divided into 32 8-dimensional capsules) and DigitCaps (fully-connected 16-dimensional capsules). 
The dynamic routing algorithm as shown in Fig.~\ref{fig:my_label-2} is used between PrimaryCaps and DigitCaps. It measures the agreement between capsules present in layer $l$ and layer $l+1$. The input to the algorithm, $u_{j/i}$, is the prediction vector for capsule $j$ in layer $l+1$ made by capsule $i$ in layer $l$. The algorithm initializes the logits $b_{ij}$ to zero and computes the coupling coefficient $c_{ij}$ for all mapping between capsule $i$ in layer $l$ and capsule $j$ in layer $l+1$. The coupling coefficients are then refined by measuring the agreement between the output $v_j$ of each capsule in layer $l+1$ and the prediction vector $u_{j/i}$. Taking the hardware perspective into consideration, these iterative computations become challenging as parallelization on a large scale is difficult.
\par Recently, the capsule networks are applied on different applications such as object segmentation~\cite{b7}, healthcare~\cite{b8} and autonomous vehicles~\cite{b9}, which are mostly implemented on GPUs. Similarly, there are other research works on capsule network configured over ASIC~\cite{b17, b18, b19}. 
Various studies have investigated FPGA-based accelerators for CNNs such as Eyeriss~\cite{b20}, AngelEye~\cite{b21}, and more recent research~\cite{b32},~\cite{b33}. However, the complex non-linear operations involved in the routing algorithm and the distinctive capsule structure make it challenging to apply these findings to capsule networks
We propose a two-step approach to deploy the CapsNet model on FPGA: 1) network pruning using our novel LAKP approach, 2) optimization of dynamic routing algorithm to reduce complexity and enable parallel operations.

\subsection{Pruning Methods}
\par Deep learning models are frequently over-parameterized and contain significant redundancies~\cite{b14}, this wastes both computation cycles and results in higher memory footprints. Network pruning can reduce the parameter redundancy and may also reduce over-fitting. There are fine-grained (unstructured) and coarse-grained (structured) pruning methods. The unstructured pruning removes the redundant parameters individually from a network based on the importance (score) of the parameters. While structured pruning removes blocks of parameters based on the cumulative score of the parameters present in the block.
Magnitude-based pruning~\cite{b22} is a simple unstructured pruning method that removes connections with the smallest weights and compresses the network efficiently. Magnitude-based pruning is a widely used pruning method due to its good compression rate and less computation cost. Park et al. presented a Look-Ahead pruning, which outperforms the magnitude-based pruning without increasing the computational cost~\cite{b15}. For pruning the $i^{th}$ layer, Look-Ahead pruning~\cite{b15} takes the weight parameters of adjacent layers $W_{i-1}$, $W_{i+1}$ in addition to the $i^{th}$ layer parameters $W_i$ and computes the look-ahead score for each parameter as given below in Equation~\ref{eq-1},   

\begin{equation}\label{eq-1}
    L_i(w)= |w| . \left\|W_{i-1}[j,:] \right\|_F . \left\|W_{i+1}[:,k] \right\|_F
\end{equation}
\par where $w$ is the weight connecting $j^{th}$ input neuron and $k^{th}$ output neuron, W[j, :] is the slice of W composed of weights connected to the $j^{th}$ output neuron, and W[:, k] is the same for the $k^{th}$ input neuron. There are other hessian-based~\cite{b23} and gradient-based~\cite{b24} pruning methods which showed better compression rate compared to magnitude-based pruning but have significantly higher computation cost.
\par Unstructured pruning shows a better compression rate compared to structured pruning but at the cost of irregularity in the sparse computation pattern. On the other hand, structured pruning, for example pruning the filter, generally will cause a lesser compression rate than pruning individual weights (as in unstructured pruning). However, in the case of larger networks containing a huge number of convolution kernels, structured kernel pruning can sometimes achieve similar or even better compression rates due to the regular pattern with a very small number of index\footnote{Represents the position of a particular weight in the original, unpruned model. It can be used to identify a specific weight within the model and to track its value during training and inference} values\cite{b14}. Structured kernel pruning, due to its structured nature, is compatible with most neural accelerators and embedded GPUs. It can also improve computational latency and reduce energy consumption by reducing the number of memory references~\cite{b14}. In another work, Sharifi et. al proposed a capsule pruning method and pruned 95$\%$ of primary capsules in the CapsNet and reduced 95.36$\%$ of floating point operations in the dynamic routing stage~\cite{b25}.
\par
\begin{figure}
    \centering
    \includegraphics[scale=0.11]{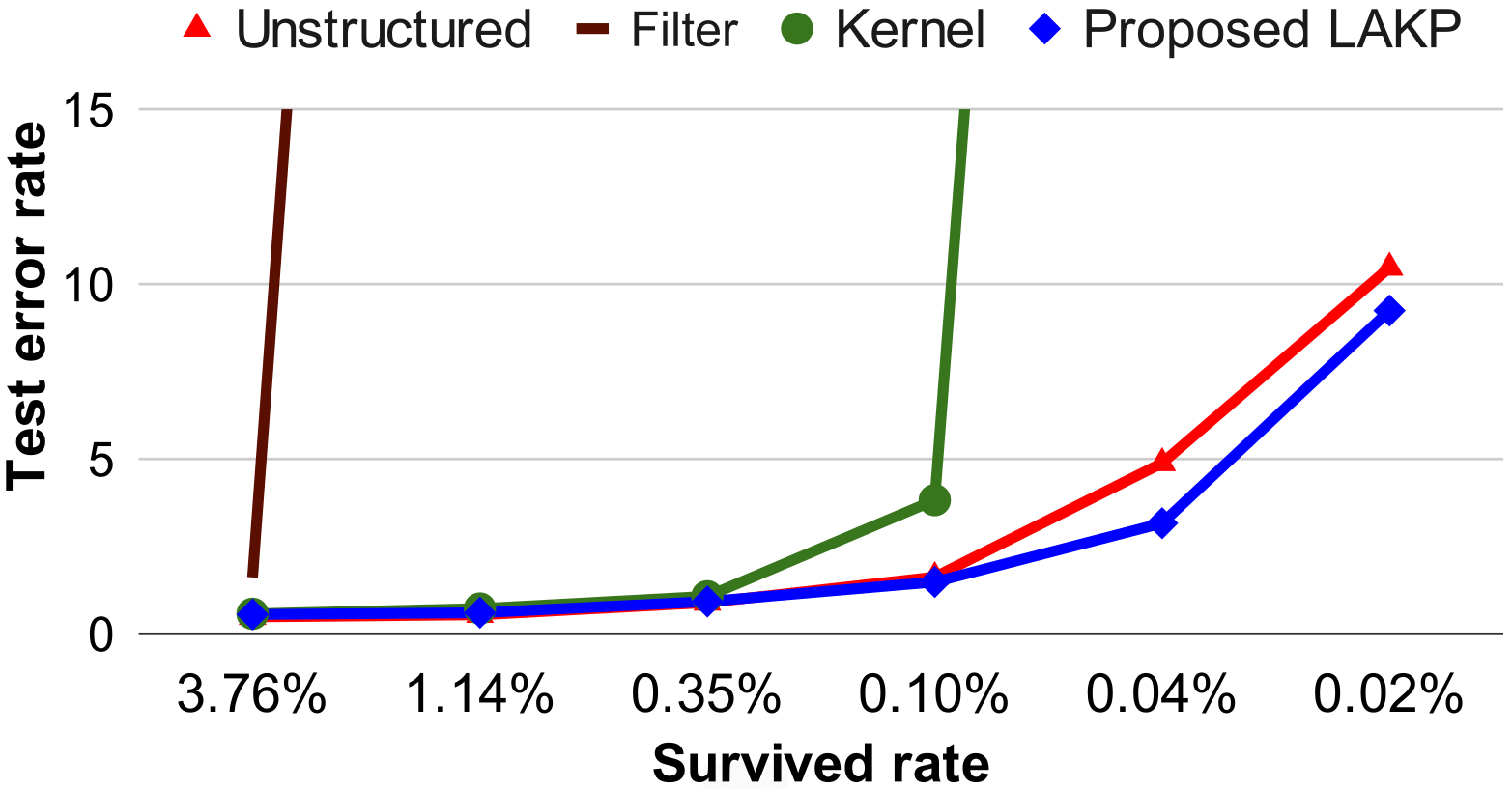}
    %\vspace{-3mm}
    \caption{Analysis of various pruning techniques on the CapsNet model utilizing MNIST data}
    \label{fig:my_label-15}
\end{figure}

 In this paper, we propose a kernel-based structured pruning method, namely Look-Ahead Kernel Pruning (LAKP), which uses the sum of look-ahead scores of the parameters and showed better compression rate compared to magnitude-based kernel pruning, particularly in the high-sparsity regime. We pruned the CapsNet using the novel LAKP method and achieved an effective compression rate of 99.26$\%$ (with an accuracy drop of less than 1$\%$) and reduced more than 99.26$\%$ of floating point operations, which is significantly better as compared to~\cite{b25}. Furthermore, the graph in Fig.~\ref{fig:my_label-15} demonstrates that the structured LAKP method represented by the blue line outperforms the unstructured magnitude based pruning method indicated by the red line in terms of compression rate when applied to the CapsNet model using MNIST data. Hence, due to the highly regular structure, the LAKP-based pruned CapsNet can be accelerated significantly faster than the unstructured magnitude-based pruned CapsNet. 

Next, we discuss our overall two-stage framework that effectively deploys a full-fledged CapsNet on FPGA.
\section{Proposed Methodology}
\par
\begin{figure}[h]
\centering
\includegraphics[scale=0.2]{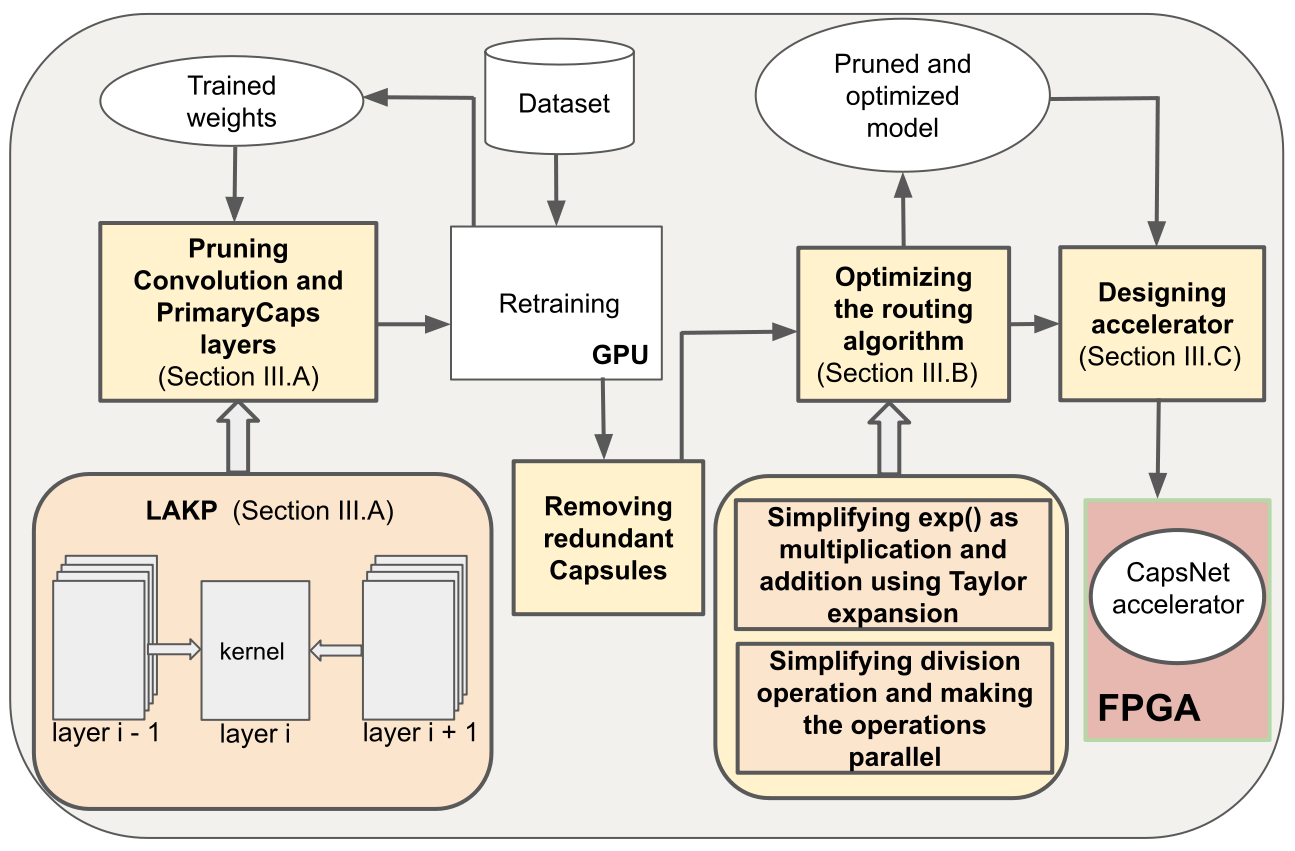}
%\vspace{-3mm}
\caption{Overview of the proposed methodology}
\label{fig:my_label-3}
\end{figure}

 In this section, we describe the implementation of CapsNet on FPGA. We propose a novel two step approach: First, we prune the model using our Look-Ahead Kernel Pruning (LAKP). Next, we optimize the routing algorithm to reduce the hardware complexity. An overview of the proposed methodology is shown in Fig.~\ref{fig:my_label-3}. The process starts with pruning the Convolution and PrimaryCaps layer kernels using the novel LAKP technique. Next, the pruned CapsNet undergoes fine-tuning to enhance the pruned model's efficiency. The interconnections between neighboring layer kernels are then studied to eliminate any unnecessary kernels and capsules from the pruned network. Finally, the dynamic routing operations are simplified to reduce hardware complexity.

Section~\ref{subsec-3a} discusses the novel LAKP method. Section~\ref{subsec-3b} discusses the proposed optimization of routing algorithm. Finally, the implementation of pruned and optimized CapsNet is described in Section~\ref{subsec-3c}.

\subsection{Look-Ahead Kernel Pruning(LAKP)}\label{subsec-3a}

\par The magnitude-based kernel pruning (KP)~\cite{b14} is a simple structured pruning technique that uses the magnitude of the parameters to identify and eliminate connections to unimportant kernels. The importance of a kernel is determined by summing the absolute values of its parameters. This method has been shown to achieve high compression rates for larger networks like VGG-19 and ResNet-18. However, it may not be suitable for all types of neural networks or data, as it does not consider the importance of a kernel to adjacent layers, which can lead to a decrease in accuracy.
We propose a Look-Ahead Kernel Pruning (LKAP), a computationally efficient algorithm to solve the optimization, which showed better compression rate compared to the state-of-the-art KP approach.  
LAKP method uses a sum of the scores for each parameter in a kernel, calculated by analyzing the effect of the parameter on the current layer and the layers adjacent to it, as the score for that kernel.
Equation~\ref{eq-1} shows the computation of the look-ahead score for each parameter. It computes the look-ahead score for a weight parameter $w$ in $i^{th}$ layer by taking the magnitude of the parameters which are present in the $i-1^{th}$ and $i+1^{th}$ layer and are directly connected to the parameter $w$.   

\par
\begin{figure}[h]
\centering
\includegraphics[scale=0.17]{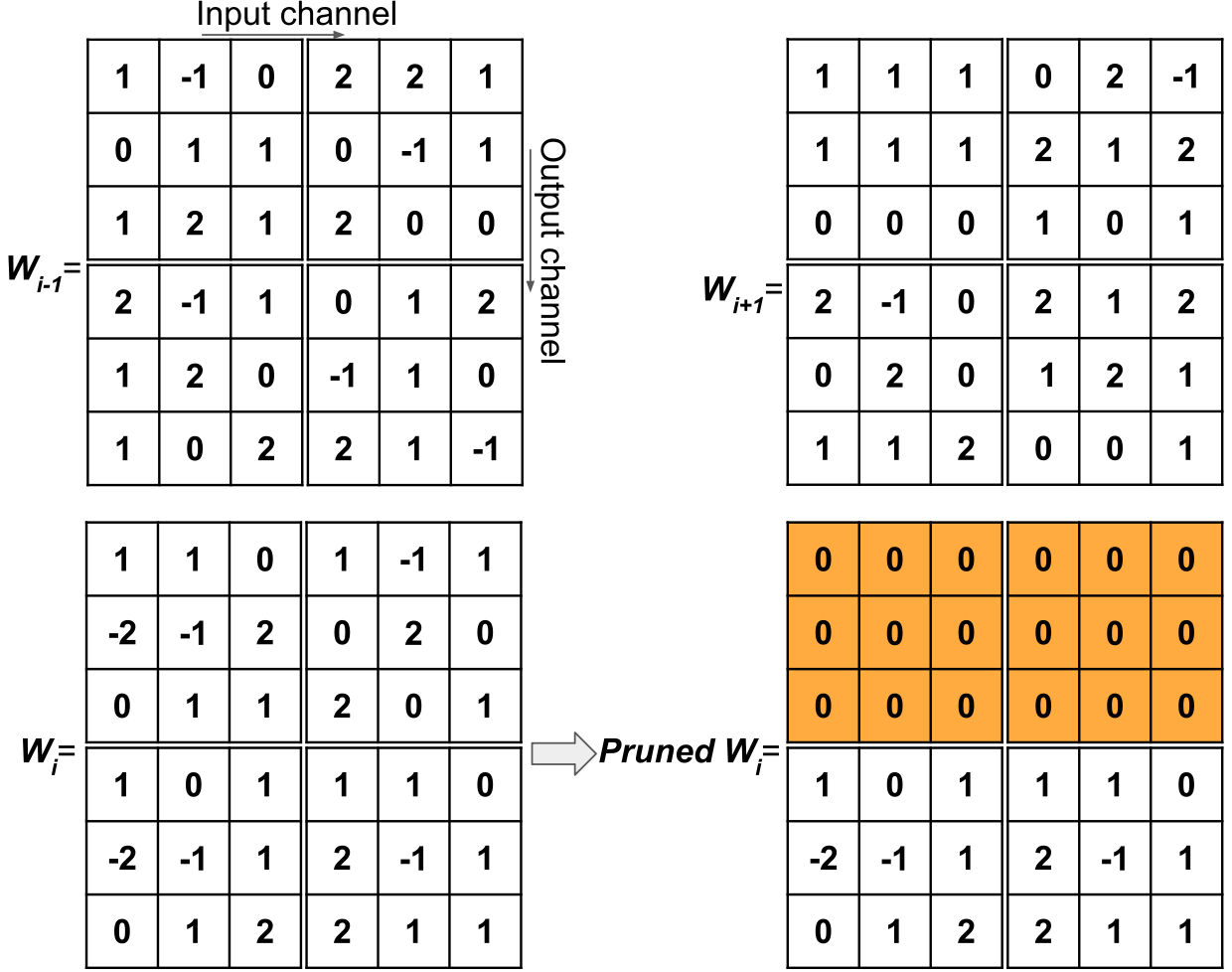}
\includegraphics[scale=0.14]{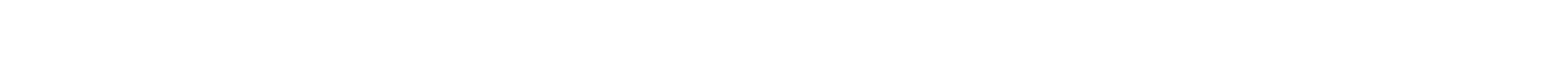}
\includegraphics[scale=0.14]{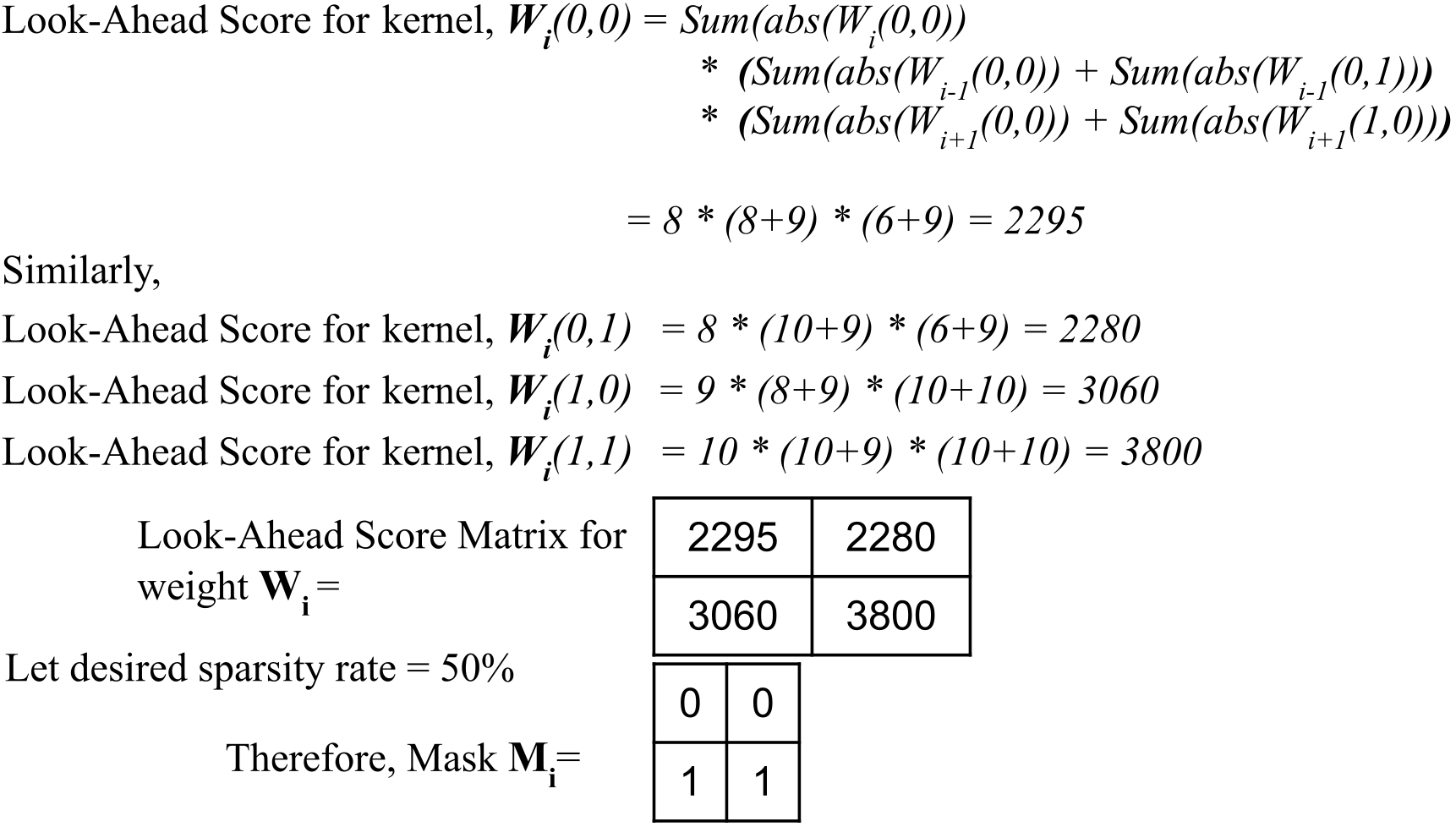}
%\vspace{-3mm}
\caption{An example to demonstrate the working of LAKP. The demonstration highlights the calculation of the look-ahead score for the kernels in weight $W_i$ of the $i^{th}$ layer and the subsequent pruning of the weight. The dimensions of $W_i$, $W_{i-1}$, and $W_{i+1}$ are presumed to be (2,2,3,3).}
\label{fig:my_label-13}
\end{figure}

 The novel LAKP is described in Algorithm~\ref{algo-1}. LAKP prunes each kernel by looking on to the importance of the parameters present in the previous and following layer that are connected to it. LAKP uses layer wise pruning method due to the unequal redundancy of network parameters in each layer~\cite{b28}. The input to the LAKP is the weight tensors of the trained network and the desired sparsity rate for each layer, then the pruned weight tensors will be given as output. Initially, the mask $M_{i}$\footnote{Mask is an $N_i$ dimensional parameter selection vector, where $N_i$ is the number of kernels in the $i^{th}$ layer} will be one for all the layers. For each layer, the algorithm computes the look-ahead score (importance) of the parameters present in that layer. The look-ahead scores of the parameters present in a particular kernel are summed and the value is taken as the score for that particular kernel. Then the masks for the parameters in the least scored kernels are zeroed. Finally, the pruned weight is the multiplication of mask and unpruned weight. The illustration of LAKP's operation can be found in Fig.~\ref{fig:my_label-13}. 

\begin{algorithm}[!ht]
\DontPrintSemicolon
  \KwInput{Weight tensors $W_1, . . . , W_L$ of a trained network, desired sparsities $s_1, . . . , s_L$}
  \KwOutput{Pruned weight tensors $\widetilde{W}_1, . . . , \widetilde{W}_L$}
  Let $N_i$ be the total number of kernels in the weight tensor $W_i$ for all $i$ $\in$ $1, ..., L$\\
  Assume each $W_i$ is flattened as $N_i$ number of kernels \\
  Initialize Mask, $M_i=\textbf{1}^{N_{i}}$ for all $i$ $\in$ $1, ..., L$\\
  \For{i = 1 to L}    
    { 
    	Compute look-ahead score $L_i(w) $ for each parameter $w \in W_i$ as shown in Eq. 1\\
    	
    	\For{j = 1 to $N_i$}    
        {
            Compute look-ahead kernel score of $j^{th}$ kernel, $LK_{j}^{i}$ := $\sum_{k_j} L_i(w)$  where $w \in k_j$ 
        }
        Set $\widetilde{k_{s_i}}$ as a $s_i$-th smallest element of  $LK^{i}$\\
        Set $M_{ij} = 0$ for all kernels, $k_j$ where $LK_{j}^{i} < \widetilde{k_{s_i}}$\\
        Set $\widetilde{W}_i := M_i . W_i$
    } 
\caption{Look-Ahead Kernel Pruning(LAKP)}
\label{algo-1}
\end{algorithm}

\par The novel LAKP method can be used to significantly reduce the size of any deep learning model that contains convolution layers. By pruning the kernels in the layer before the capsule layer in the case of the CapsNet, it leads to a decrease in the number of capsules in the capsule layer, resulting in a substantial enhancement in the network's performance as the capsule layer comprises of complex calculations. Experiments on MNIST and F-MNIST datasets showed that the proposed LAKP nullified the number of capsules in the PrimaryCaps layer from 1152 to 252 and 432 respectively, resulting in a reduction of weight parameters in the capsule layer by 1280 times as each capsule operates with 10 * 16 * 8 weight parameters.

\subsection{Optimizing routing algorithm}\label{subsec-3b}
The routing algorithm contains frequent non-linear operations such as exponential (exp()) and division (div()) which consumes more clock cycles compared to other operations. Simplifying these operations can reduce the latency of the computation. By using the Taylor expansion, an approximation of the exponential function can be written as shown in Equation~\ref{eq-2}, 
\begin{dmath}\label{eq-2}
e^x=e^a ( 0.60653 + x ( 0.60659 + x ( 0.30260 + x ( 0.10347 + x ( 0.02118 + 0.00833 x )))))
\end{dmath}

where $a=0.5$. Multiplying the $e^a$ term prior and taking only the first 5 components, the operation can be executed using just 5 multiplications and 5 additions without dropping accuracy. Thus the latency of exp() is reduced from 27 cycles to 14 cycles.   

\par The fixed point division operation also requires a latency of 49 cycles which is very high as compared to all other operations. Hence, we optimized the division operation and represented it with exponential and logarithm operations as given below in Equation~\ref{eq-3},
\begin{equation}\label{eq-3}
    a/b= e^{log(a)-log(b)}
\end{equation}
where $a$ and $b$ are arbitrary real numbers. This reduced the latency of fixed point division operation from 49 cycles to 36 cycles. Moreover, the exponential operation is implemented using an array of Processing Element (PE) containing element wise multiplication and adder tree. We implemented an array of 10 PEs which improved the throughput of the CapsNet model trained on MNIST by \textbf{615} FPS. Since the exp() and div() are used frequently in Softmax() operation ($4^{th}$ step in the routing algorithm shown in Fig.~\ref{fig:my_label-2}), the operation is simplified by the proposed optimization method. 

\par
\begin{figure}[h]
    \centering
    \includegraphics[scale=0.15]{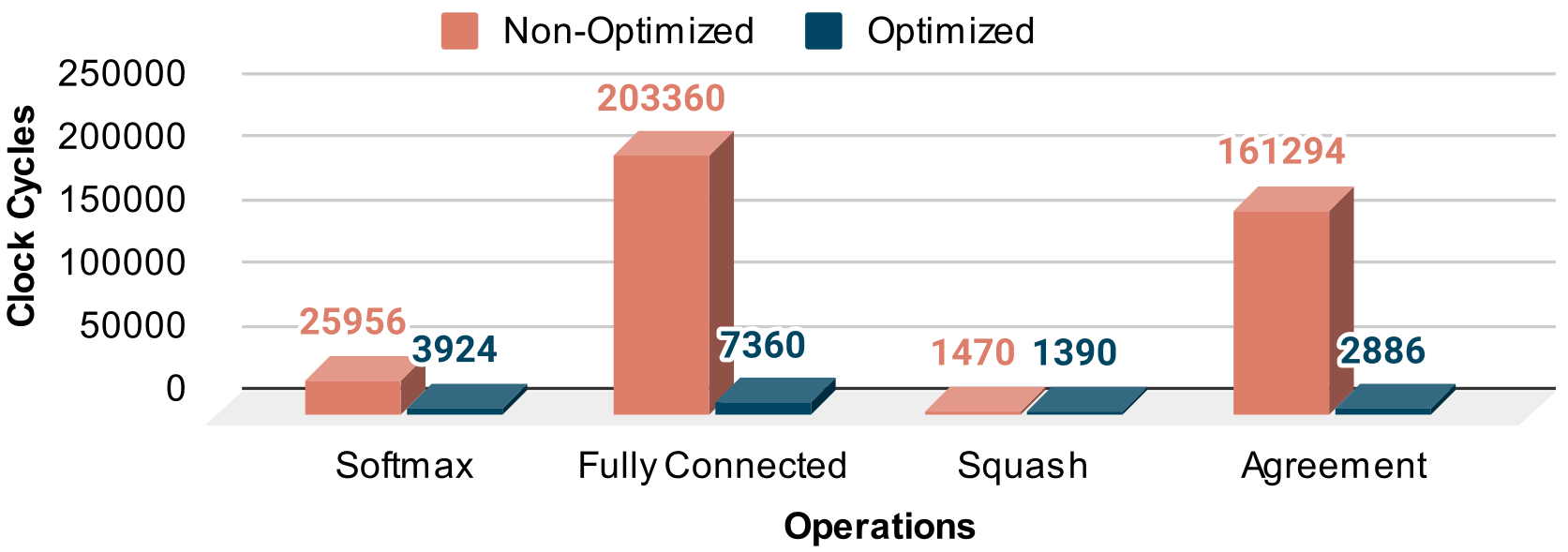}
    %\vspace{-3mm}
    \caption{Latency of each operation in the dynamic routing algorithm of pruned CapsNet model trained on MNIST}
    \label{fig:my_label-4}
\end{figure}

 Additionally, we have accelerated the Agreement and the Fully Connected steps (in the routing algorithm, see Fig.~\ref{fig:my_label-2}) by making operations parallel. We made the operations parallel by reordering the loops and utilizing the Vivado HLS
directives and PEs. 
Code 1 shows the pseudo code for the Agreement step of the routing algorithm. To reduce the writing conflict and increase parallelism we reordered the loops. Code 2 shows the reordered loops for the Agreement step of the routing algorithm. The number of clock cycles required for each operation in the optimized and non-optimized routing algorithm is given in Fig.~\ref{fig:my_label-4}. The diagram displays a substantial decrease in the number of clock cycles needed per routing operation. These routing operations will be repeatedly carried out, resulting in a substantial increase in both throughput and energy efficiency.
% \begin{itemize}
% \end{itemize}
\fbox{\begin{minipage}{25em}
\For{i= 1 to $IN\_CH$}    
    { 
        \For{j=1 to $OUT\_CH$}
        {
            \For{k=1 to $OUT\_DIM$}
            {
                b[i][j] += u[i][j][k] * v[j][k]
            }
        }
    }
% \caption{Code 1: Pseudo code for the Agreement step}

\end{minipage}}
\vspace{0.2cm}
% \begin{itemize}
% \end{itemize}
\fbox{\begin{minipage}{25em}
\For{j= 1 to $OUT\_CH$}    
    { 
        \For{k=1 to $OUT\_DIM$}
        {
            \For{i=1 to $IN\_CH$/ fact}
            {
                \#pragma HLS PIPELINE II=1\\
                PE( b[i*fact : i*fact+fact][j] , u[i*fact : i*fact+fact][i][j] , v[j][k] )
            }
        }
    }
% \caption{Code 2: Reordered loops for the Agreement step}
\end{minipage}}

%\par Furthermore, We can reduce the latency of exp() and log() operations further by logically defining them in a hardware description language (HDL), in our case Verilog. The work by~\cite{b27} showed that we can reduce the latency of exp() and log() to nearly 1 to 3 cycles. The Vivado HLS (we used Vivado HLS for creating IPs) enables the use of existing Verilog RTL IP in an HLS project. We will utilize this RTL black box feature for our future work.
%\squeezeup
\subsection{Deploying pruned model on FPGA}\label{subsec-3c}
\par
\begin{figure}[h]
    \centering
    \includegraphics[scale=0.17]{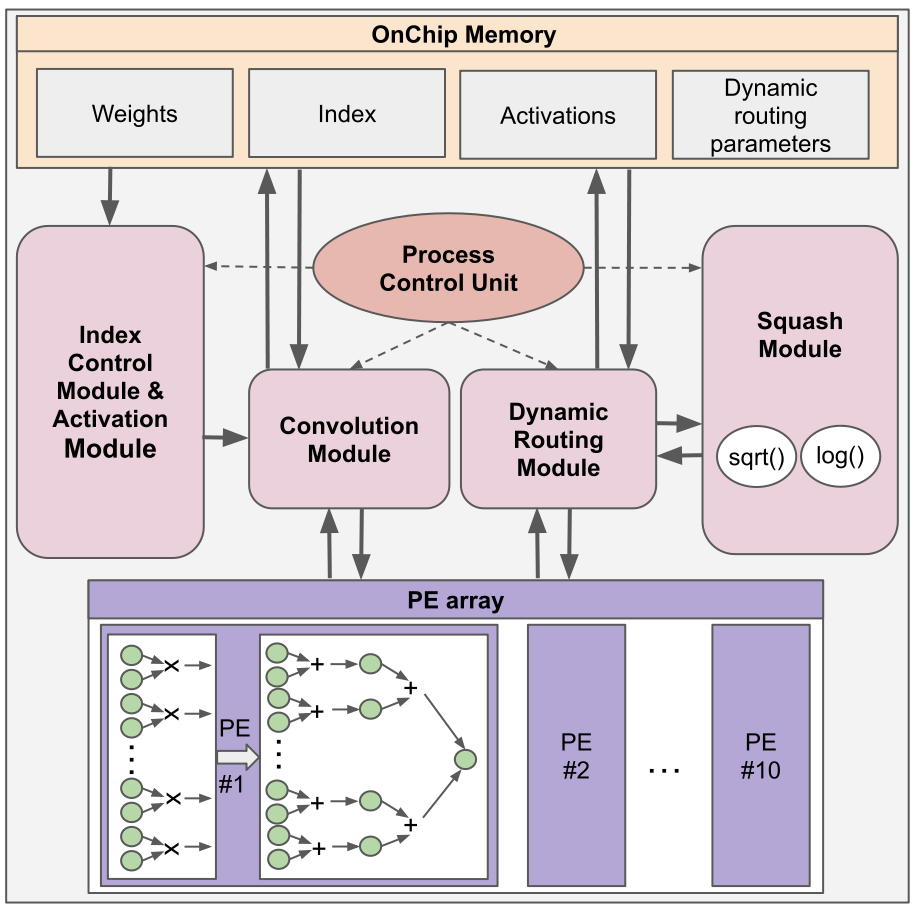}
    %\vspace{-3mm}
    \caption{Architecture of the proposed CapsNet accelerator}
    \label{fig:my_label-5}
\end{figure}

 When implementing a pruned network on an FPGA, extra memory units are needed to keep the indices of the weights that were not pruned. With the help of our proposed structured kernel pruning, we only retain the index of the kernels that haven't been pruned. This is in contrast to each weight index as in case of unstructured pruning. This allows us to use minimal memory to store the indices, which is only 0.1$\%$ of the total number of weights that remain after pruning. In order to reduce data transfer overhead, all the parameters are  saved on-chip. Fig.~\ref{fig:my_label-5} depicts the design of the proposed CapsNet accelerator.

\begin{figure}[h]
    \centering
  
  \subfloat[\label{1a}]{%
        \includegraphics[scale=0.12]{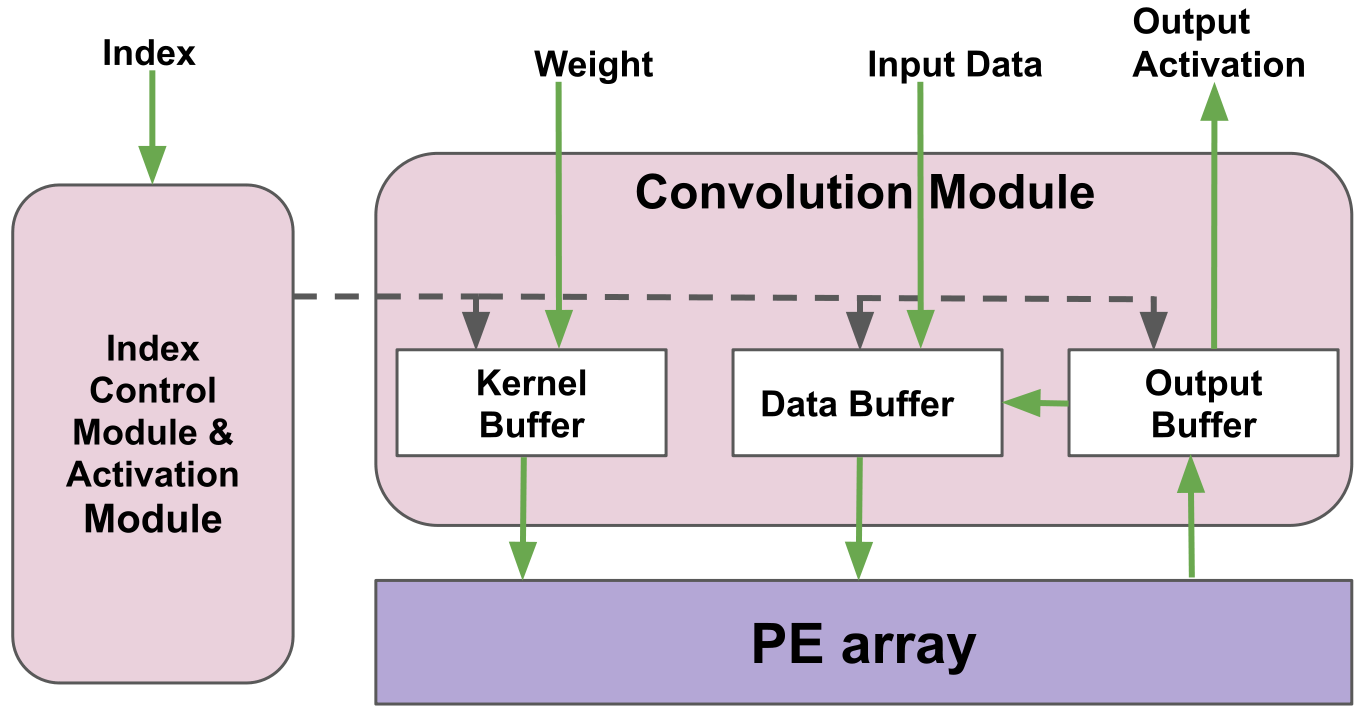}}
    \\ 
    
  \subfloat[\label{1b}]{%
         \includegraphics[scale=0.16]{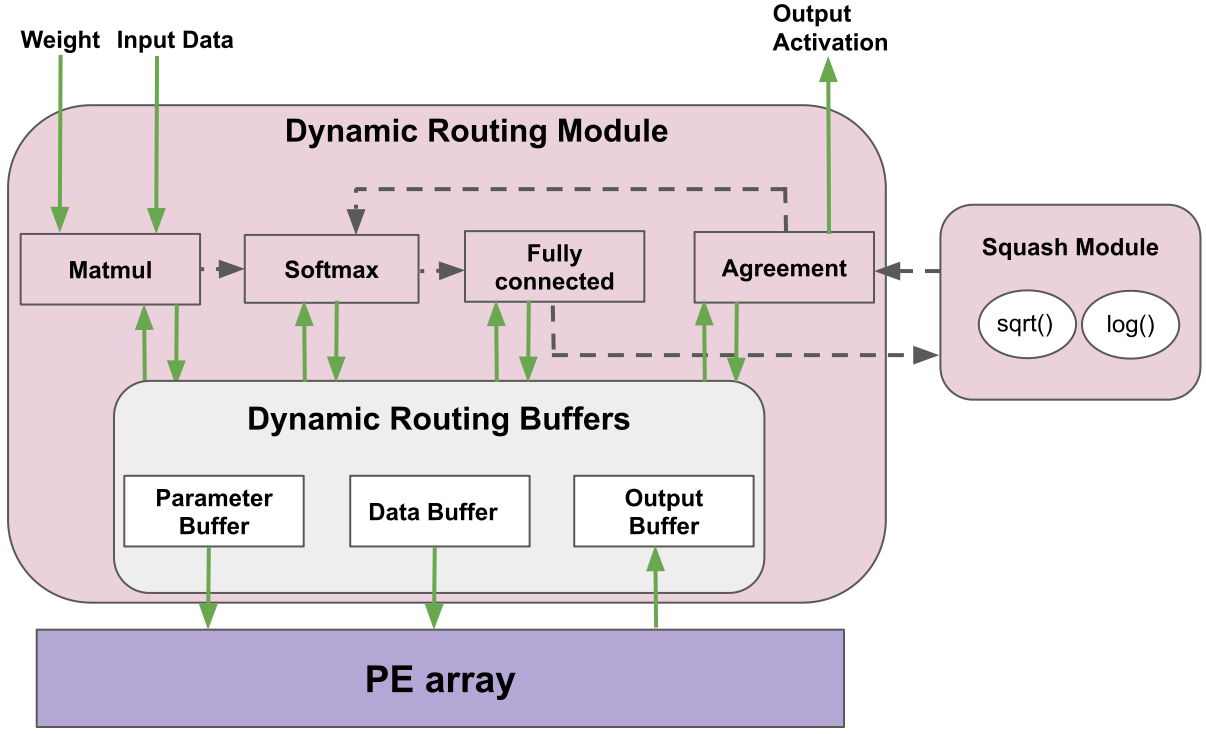}}
  %\vspace{-3mm}
  \caption{Demonstrates the data flow of (a) Convolution Module and (b) Dynamic Routing Module}
  \label{fig:my_label-9} 
\end{figure}

\begin{figure}[h]
    \centering
  
  \subfloat[\label{1a}]{%
        \includegraphics[scale=0.14]{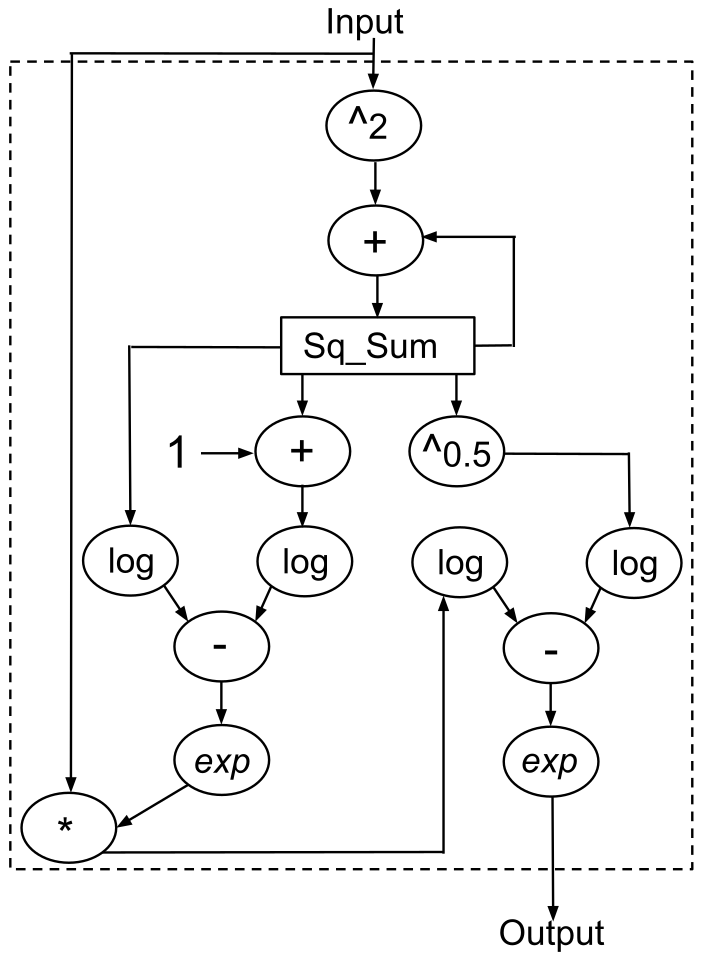}}  \hspace{0.2cm}
        \subfloat[\label{1b}]{%
        \includegraphics[scale=0.15]{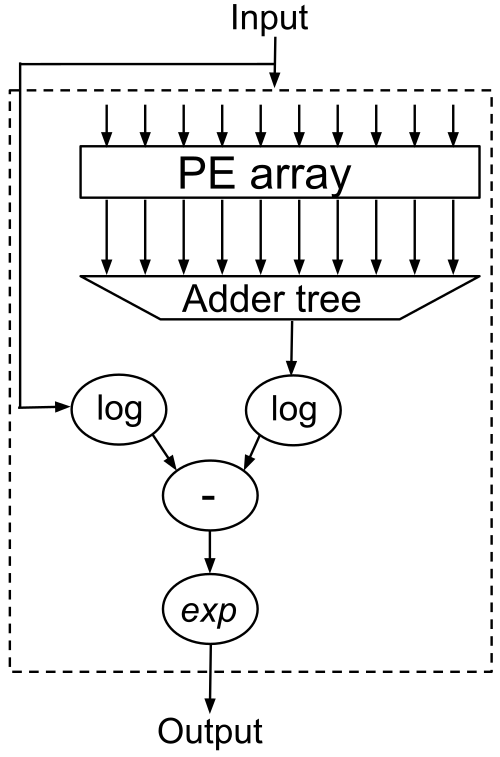}}
  %\vspace{-2mm}  
  \caption{Architecture of (a) Squash Function and (b) Softmax Function}
  \label{fig:my_label-14} 
\end{figure}

It uses on-chip memory to store weights, activation outputs, index values for survived weights, and dynamic routing parameters. The convolution and dynamic routing modules handle the memory transactions and the step-by-step procedures involved in the convolution and dynamic routing algorithms, respectively. The Convolution Module (Fig.~\ref{fig:my_label-9}(a)) retrieves the weight kernel and input data from the on-chip memory, the Index Control Module maps the kernel and input data based on the index, and the computation is performed on the PE array. The resulting output activations are indexed and stored in the output memory. Similarly, the Dynamic Routing Module (Fig.~\ref{fig:my_label-9}(b)) obtains the routing parameters and input data from the on-chip memory and performs the computation using the PE array. To improve throughput and resource utilization, all routing steps except the Squash operation are executed on the PE array. Fig.~\ref{fig:my_label-14} illustrates the comprehensive architecture of the Squash and Softmax functions. The index control module helps make the MAC operations parallel and store the results efficiently so that the following layer can utilize it. An array of 10 PEs is implemented that performs element-wise 16-bit nine multiplications followed by an adder tree which adds up all the nine multiplication results. The Process Control Unit manages the overall data flow of the proposed pruned CapsNet accelerator.
\par The exp() function is simplified to multiplications and additions, the operation is executed using the array of PE's. The latency of softmax() operation is reduced by $85\%$ in the unpruned and pruned models, improving the throughput of the pruned CapsNet model trained on MNIST by 615 FPS. The optimization strategy discussed in Section III. B can be used for any capsule network without changing the design entirely.  

\section{Experimental Results}
\begin{figure}[h]
\centering
\includegraphics[scale=0.148]{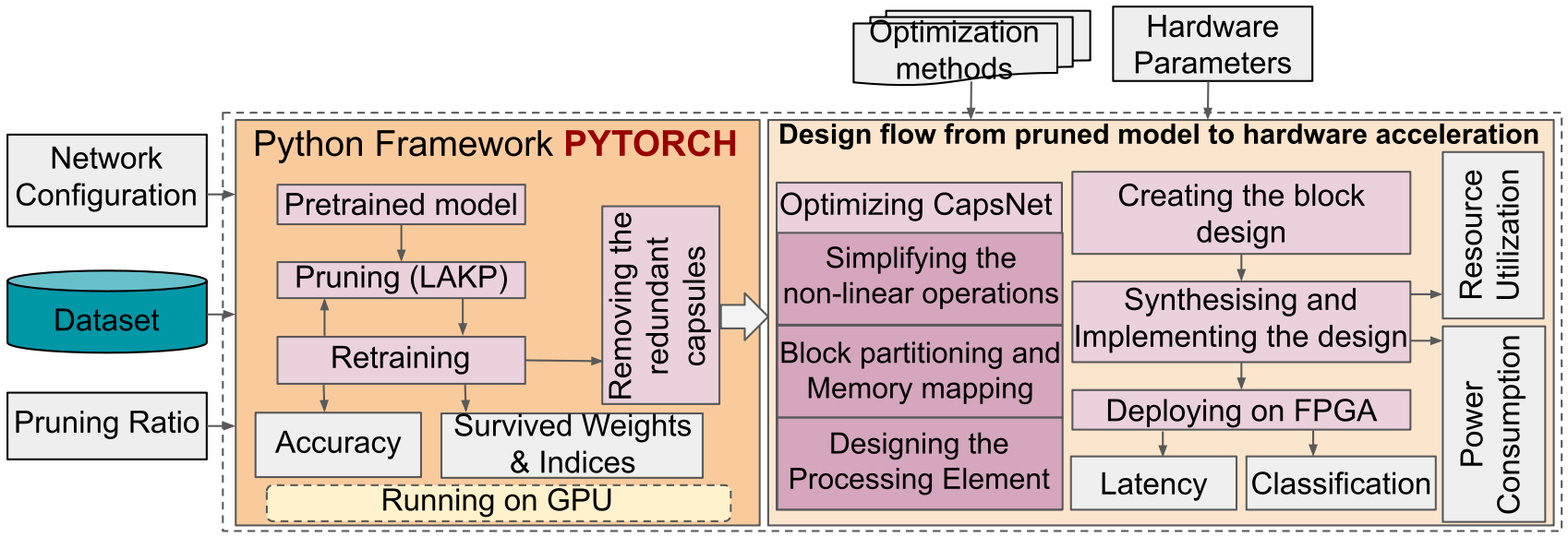}
%\vspace{-3mm}
\caption{The experimental setup and implementation}
\label{fig:my_label-11}
\end{figure}

In this section, we present a detailed discussion on our experimental setup shown in Fig.~\ref{fig:my_label-11} for various performance analysis. We first validate the effectiveness of our novel pruning algorithm LAKP compared to the state-of-the-art~\cite{b14}. Next, we discuss the deployment and comparative performance analysis of the original CapsNet with our proposed pruned and optimized CapsNet. 

\subsection{Evaluating LAKP over KP }\label{subsec-4a}
\par We used Google Colab with GPU accelerator for training the models. We compared the empirical performance of LAKP with KP. We looked at three different neural network architectures: CapsNet\cite{b4}, VGG-19\cite{b29} and ResNet-18\cite{b30}. In particular, CapsNet is trained on MNIST and F-MNIST, while VGG-19 and ResNet-18 are trained on CIFAR-10 and GTSRB. Table~\ref{table-1} displays the test error rates of the pruned models corresponding to the percentage of the survived weights. The last entry in the first row of Table I, 0.6, represents the test error rate of the CapsNet model, trained on MNIST and pruned using LAKP at a pruning rate of 98.86$\%$ (100-1.14). The table shows that LAKP regularly outperforms KP in the aforementioned models, especially when there is significant sparsity, with over 89$\%$ gain on CapsNet and 34$\%$ on VGG-19.

\begin{table}[!ht]
\caption{Test error rates of the models. Bracketed
numbers denote relative gains with respect to KP}
\centering
\begin{tabular}{p{0.11\linewidth}p{0.11\linewidth}p{0.12\linewidth}p{0.10\linewidth}p{0.07\linewidth}p{0.22\linewidth}}
\hline
Model & Dataset & Actual test error & Survived rate & Test error (KP) & Test error (proposed-LAKP) \\
\hline
 &  &   & 1.14$\%$ & 0.72 & \textbf{0.6} ($-16.7\%$)\\

& & & 0.35$\%$ &  1.07 &  \textbf{0.92} ($-14.0\%$) \\
CapsNet&MNIST & 0.67& 0.1$\%$ &  3.82 & \textbf{1.48} ($-61.3\%$) \\

& & & 0.04$\%$ & 88.65 & \textbf{3.16} ($-96.4\%$) \\
 
& & & 0.02$\%$  &  88.65 & \textbf{9.24} ($-89.6\%$)\\
\hline
 & & & 17.88$\%$ & 9.63 & \textbf{9.24} ($-4.0\%$)\\

CapsNet&F-  & 10.31 & 5.69$\%$ &  10.67 &  \textbf{10.03} ($-5.9\%$) \\
&MNIST & & 1.37$\%$ &  12.58 & \textbf{11.82} ($-6.0\%$) \\

& & & 0.25$\%$  & 17.64 & \textbf{15.04} ($-14.7\%$) \\
 
& & & 0.01$\%$   &  89.65 &  \textbf{32.50} ($-63.7\%$)\\
\hline
& &   & 21.59$\%$ & 9.18 & \textbf{8.98} ($-2.2\%$)\\
VGG-&CIFAR- & 9.09 & 10.26$\%$ & 10.93 &  \textbf{10.74} ($-1.7 \%$) \\
19& 10&  & 4.31$\%$ &  16.7 & \textbf{15.49} ($-7.3\%$) \\

& & & 3.82$\%$  & 21.86 & \textbf{16.12} ($-26.3\%$) \\
 
& & & 1.93$\%$   &  51.27 &   \textbf{33.72} ($-34.2\%$)\\
\hline
 &  &  & 13.63$\%$ & 0.54 & \textbf{0.46} ($-14.8\%$)\\
VGG-& GTSRB& 1.15& 7.39$\%$ & 0.54 &  \textbf{0.36} ($-33.3 \%$) \\
19& &   & 3.82$\%$ &  1.1 &\textbf{0.54} ($-50.9\%$)\\
& & & 1.93$\%$  & 92.42 & \textbf{1.91} ($-97.9\%$) \\

& & & 0.96$\% $  &  99.13 &  \textbf{60.64} ($-38.8\%$)\\
\hline
 &  &   & 21.8$\%$ &  8.59 & \textbf{8.26} ($-3.8\%$)\\
ResNet-&CIFAR-  &8.83 & 12.03$\%$ & 9.51 &  \textbf{9.16} ($-3.7\%$)\\
18&10 & & 6.11$\%$ &  11.24 &\textbf{11.03} ($-1.9\%$)\\

& & & 2.89$\%$  & 16.79 &  \textbf{15.87} ($-5.5\%$) \\
 
& & & 1.3$\%$  &  24.29 &  \textbf{20.77} ($-14.5\%$)\\
\hline
&  &  & 1.97$\%$ &   0.69  & \textbf{0.36} ($-47.8\%$)\\
ResNet-&GTSRB & 0.54 & 0.73$\%$ &2.76 &  \textbf{2.30} ($-16.7\%$)\\
18& & & 0.26$\%$  &  8.90 & \textbf{6.51} ($-26.9\%$)\\
& & & 0.09$\%$  & 23.57 &  \textbf{17.40} ($-26.2\%$) \\
& & & 0.04$\%$  &  29.97 &  \textbf{29.06} ($-3.0\%$)\\
\hline
\end{tabular}
\label{table-1}
\end{table}

\subsection{Deploying the model on FPGA}\label{subsec-4b}
\par
\begin{figure}[h]
\centering
\includegraphics[scale=0.08]{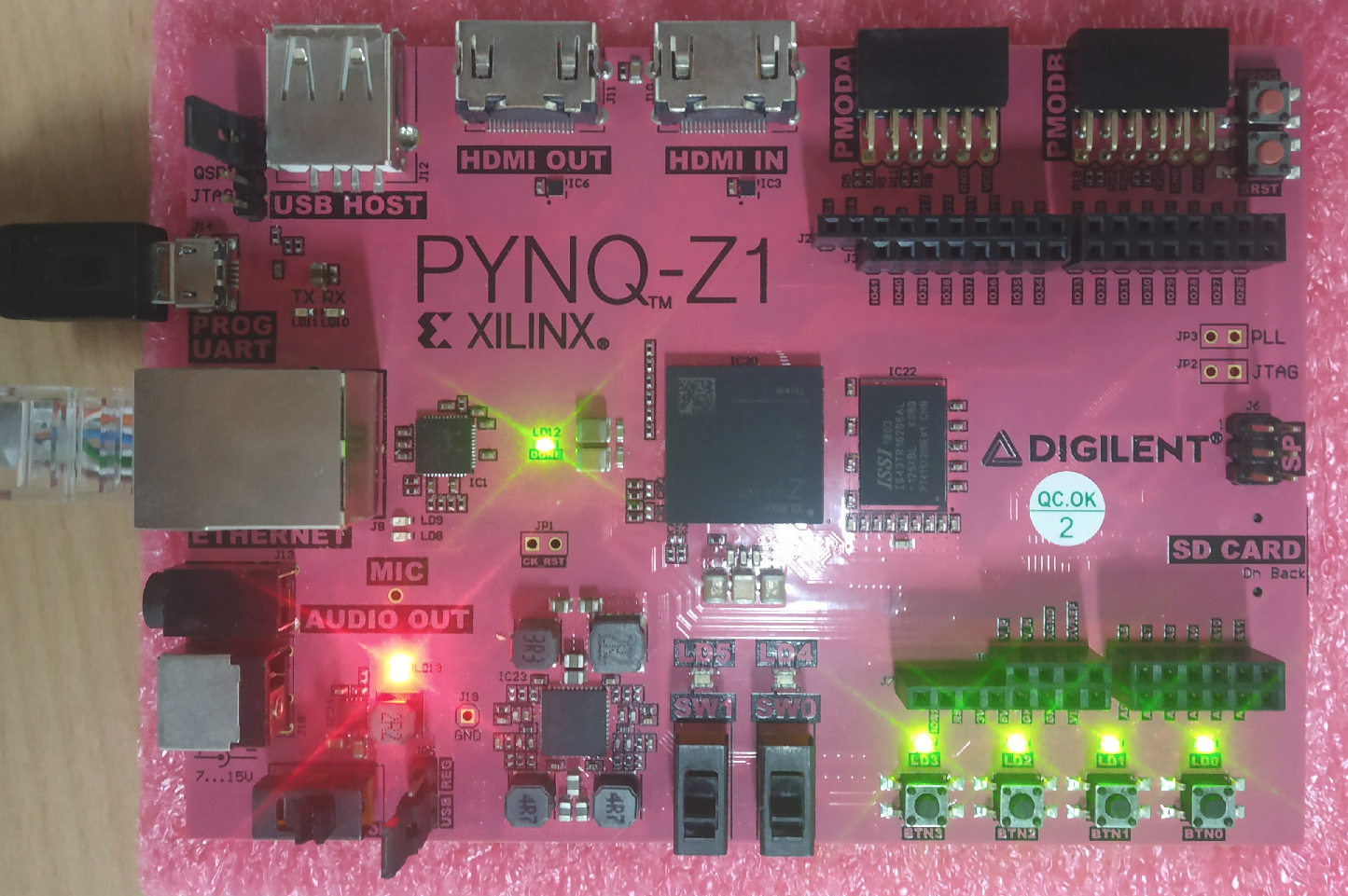}
%\vspace{-3mm}
\caption{Xilinx PYNQ-Z1 FPGA board used in  our experiments and prototype}
\label{fig:my_label-12}
\end{figure}

\begin{figure}[h]
\centering
\includegraphics[scale=0.15]{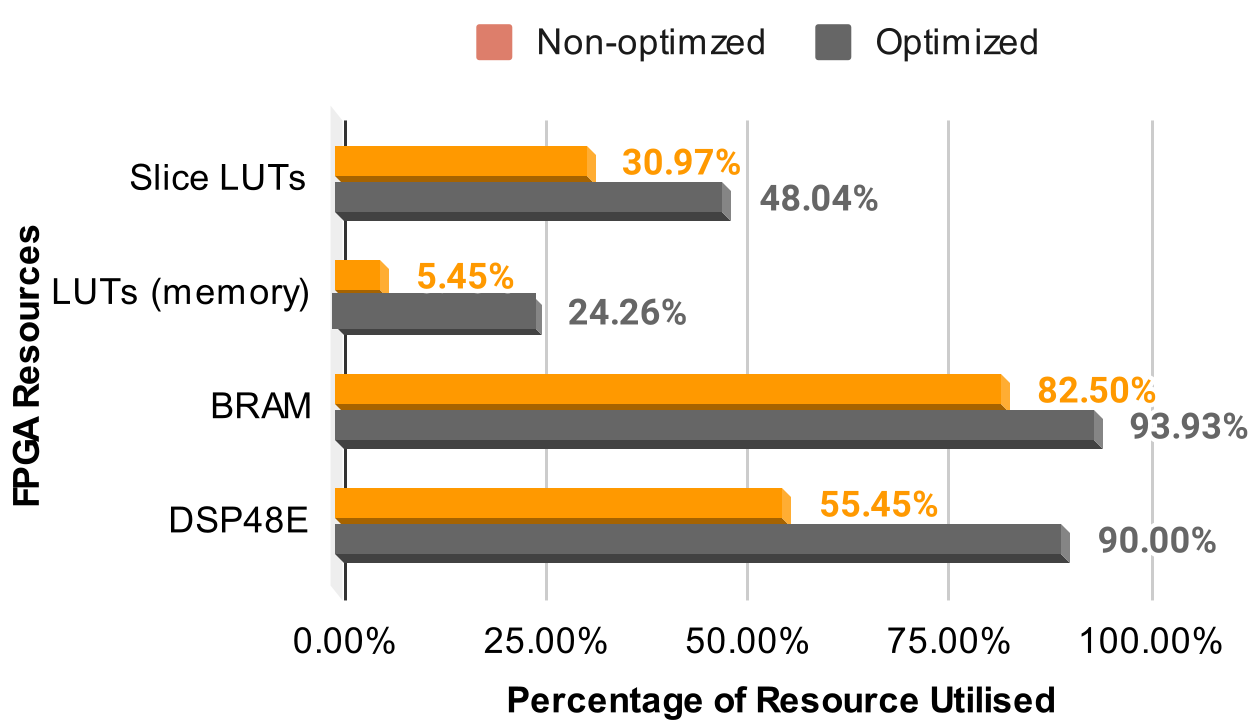}
%\vspace{-3mm}
\caption{Comparison of resource utilization of Non-optimized and Optimized pruned CapsNet (trained on MNIST)}
\label{fig:my_label-6}
\end{figure}

 We used Xilinx PYNQ-Z1 FPGA board for the experiments, as depicted in Fig.~\ref{fig:my_label-12}. In the first set of experiments we deployed the original CapsNet model shown in Fig.~\ref{fig:my_label-1} and our proposed pruned and optimized CapsNet on FPGA. The number of parameters in original CapsNet model limits the usage of Vivado HLS optimization directives due to the excessive usage of available resources. Since the original model contains a large number of operations, it delivers very low performance of 5 FPS.
After applying LAKP method to the original model, we observed a performance of 82 FPS, which got further improved to 1351 FPS after applying mathematical optimization of routing algorithm functions. Fig.~\ref{fig:my_label-6} compares the resource utilization of Non-optimzed and Optimized pruned CapsNet model. The optimization method makes non-linear operations more hardware-friendly and able to be parallelized, thus it does not consume a lot of resources. Therefore, the proposed optimization method can be applied to any CapsNet model being deployed on resource-constrained hardware such as ASIC and FPGA. In Table~\ref{table-2}, we have presented the recorded difference in the resource utilization between the original CapsNet model and our proposed (pruned $+$ optimized) CapsNet.

Next, we trained the CapsNet using the F-MNIST dataset and observed that the frame rate of the pruned model is improved from 48 FPS (non-optimized routing algorithm) to 934 FPS (optimized routing algorithm). In Table~\ref{table-3}, resource utilization of the pruned and optimized CapNet trained on F-MNIST dataset is shown, that again validates the effectives of our proposed novel LAKP approach and dynamic routing optimization framework. Furthermore, we implemented 16-bit quantization to the network parameters, and the proposed optimization approach did not lead to a reduction in the accuracy of the network.  

\begin{table}[!ht]
        \caption{Resource utilization Original CapsNet\cite{b4} and Proposed CapsNet on FPGA (trained on MNIST)}
        \centering
        \begin{tabular}{p{0.25\linewidth}p{0.3\linewidth}p{0.25\linewidth}}
            \hline
            Resources &  Original CapsNet\cite{b4} & Proposed CapsNet \\
        
            \hline
            Slice LUTs & 33232 (62.47 $\%$) & 25559 (48.04 $\%$)\\
            LUTs (memory) & 6751 (38.80 $\%$) & 4221 (24.26 $\%$)\\
            BRAM & 140 (100 $\%$) & 131.5 (93.93 $\%$)\\
            DSP48E & 187 (85.00 $\%$) & 198 (90 $\%$) \\
            \hline
            Latency(1 sample) & 0.19 sec & 0.00074 sec\\
            \hline
        
        \end{tabular}
\label{table-2}
\end{table} 

\begin{table}[!ht]
        \caption{Resource utilization of Proposed CapsNet(trained on F-MNIST)}
        \centering
        \begin{tabular}{p{0.25\linewidth}p{0.25\linewidth}}
            \hline
            Resources &  Proposed CapsNet \\
            
            \hline
            Slice LUTs  & 28247 (43.10 $\%$)\\
            LUTs as memory  & 6268 (36.02 $\%$)\\
            BRAM & 131.5 (93.93 $\%$)\\
            DSP48E & 198 (90 $\%$) \\
            \hline
            Latency(1 sample) & 0.00107 sec\\
            \hline
        \end{tabular}
\label{table-3}
\end{table}

\section{Conclusion}
In this paper, we propose a novel two-step approach to deploy the CapsNet model on FPGA: First, the network is pruned using our novel LookAhead Kernel Pruning (LAKP) approach. Then, the dynamic routing algorithm is optimized to reduce hardware complexity and make operations parallel. Experiments on the models CapsNet, VGG-19 and ResNet-18 show that the proposed LAKP consistently perform better compared to the state-of-the-art magnitude-based kernel pruning. Experiment on CapsNet with MNIST and F-MNIST datasets, the proposed LAKP achieved an effective compression rate of $99.26\%$ and $98.84\%$ with an accuracy drop of less than 1$\%$. As a result of optimization of the routing algorithm, the pruned CapsNet models trained on MNIST and F-MNIST achieved a throughput of 1351 FPS and 934 FPS respectively over Xilinx PYNQ-Z1 FPGA. As corroborated by our results, this work enables highly performance-efficient deployment of CapsNets on low-cost FPGA.   

\section*{Acknowledgment}

This work was supported in parts by the NYUAD's Research Enhancement Fund (REF) Award on "eDLAuto: An Automated Framework for Energy-Efficient Embedded Deep Learning in Autonomous Systems", and by the NYUAD Center for Artificial Intelligence and Robotics (CAIR), funded by Tamkeen under the NYUAD Research Institute Award CG010.
    
\balance

\vspace{12pt}

\end{document}